\documentclass[12pt]{iopart}

\usepackage{iopams}  

\usepackage{amsthm,graphicx}

\usepackage{mathrsfs}
\usepackage{amssymb}

\def\be{\begin{equation}}
\def\ee{\end{equation}}
\def\bea{\begin{eqnarray}}
\def\eea{\end{eqnarray}}
\def\bean{\begin{eqnarray*}}
\def\eean{\end{eqnarray*}}

\def\scri{\mathscr{J}}

\begin{document}

\note[Black hole formation by electromagnetic radiation]{Black hole formation 
by incoming electromagnetic radiation}

\author{Jos\'e M M Senovilla$^1$}

\address{$^1$ F\'{\i}sica Te\'orica, Universidad del Pa\'{\i}s Vasco, Apartado 644, 48080 Bilbao, Spain}
\eads{\mailto{josemm.senovilla@ehu.es}}
\begin{abstract}
I revisit a known solution of the Einstein field equations to show that it describes the formation of non-spherical black holes  by the collapse of pure electromagnetic monochromatic radiation. Both positive and negative masses are feasible without ever violating the dominant energy condition. The solution can also be used to model the destruction of naked singularities and the evaporation of white holes by emission or reception of light.
\end{abstract}

\pacs{04.70.Bw, 04.40.Nr, 04.20.Jb}

Today there are many models describing the formation, or the evaporation, of black holes in General Relativity, some of them deal with the collapse of matter or fluids, others with out- or in-coming ``incoherent radiation''. However, there is no identified case of the formation/evaporation of a black hole by means of electromagnetic radiation solely. The purpose of this sort Note is to call attention to a family of solutions which describe the formation/destruction of black holes (also naked singularities) by reception or emission of pure monochromatic light. Actually, the solution is known since long ago, and has been used to describe the collapse of null dust to non-spherical black holes with negative cosmological constant \cite{L,L2}, and to  analyze the first example of a dynamical horizon with toroidal topology \cite{DFM}. However, the fundamental point is that the radiation can be properly identified as monochromatic light ---it should not be treated as ``incoherent radiation''.

The line-element of the solution was first presented ---among many other exact solutions--- by Robinson and Trautman in their celebrated paper \cite{RT}, and is given in local coordinates $\{u,r,x,y\}$ by (see also \cite{Exact}, p.430)
\be
ds^2=r^2(dx^2+dy^2) +2\epsilon dvdr +\left(\frac{2m(v)}{r}+\frac{\Lambda}{3} r^2 \right)dv^2  \label{ds2}
\ee
where $r>0$,  $\epsilon=\pm 1$ determines the retarded ($\epsilon =-1$) or advanced ($\epsilon =1$) character of the null coordinate $v$, $\mathbf{k}=-dv$  is a null one-form chosen to be {\em future} pointing, $\Lambda$ is the cosmological constant (allowed to have any sign) and $m(v)$ is a function of $v$.

The metric (\ref{ds2}) is a solution of the Einstein-Maxwell equations with $\Lambda$ for a {\em null electromagnetic field} given by
\be
\mathbf{F}=dv \wedge (h_x(v)dx +h_y(v) dy) \label{F}
\ee
so that $\mathbf{k}$ is the wave vector of this pure radiation field. Here, $h_x(v)$ and $h_y(v)$ are arbitrary functions. An appropriate observer with unit timelike tangent vector $\vec u$ orthogonal to the $\{x,y\}$ surfaces and with $u^\mu k_\mu =-1$ will measure electric $\mathbf{E}$ and a magnetic $\mathbf{B}$ fields given by
$$
\mathbf{E} = h_x(v)dx +h_y(v) dy, \hspace{1cm} \mathbf{B}= h_y(v)dx -h_x(v) dy \, .
$$
Therefore, by choosing the functions $h_x,h_y$ judiciously one can describe light that is linearly polarized (e.g. $h_y=0$), circularly polarized (e.g. $h_x=A \cos v$, $h_y =A \sin v$), or elliptically polarized.

The energy-momentum tensor of the electromagnetic radiation reads
\be
T_{\mu\nu}= \frac{h_x^2(v)+h_y^2(v)}{r^2} k_\mu k_\nu \label{emt}
\ee
from where one determines the function $m(v)$ on the line-element (\ref{ds2}):
\be
m(v) =\epsilon \frac{4\pi G}{c^4}  \int^v_{v_0} [h_x^2(w)+h_y^2(w)]dw + M \label{mass}
\ee
where $v_0$ is a fixed value of $v$ and $M=m(v_0)$ is an integration constant ($v_0$ can be $-\infty$). The energy-momentum tensor (\ref{emt}) always satisfies the dominant energy condition, and we have
$$
\dot m (v) =\epsilon \frac{4\pi G}{c^4} [h_x^2(v)+h_y^2(v)]
$$
where dots denote derivatives with respect to $v$, so that the function $m(v)$ is non-decreasing for $\epsilon =1$ and non-increasing for $\epsilon =-1$.
Observe that $M$ can have any sign.

The metric (\ref{ds2}) is the unique Petrov type D solution of the Einstein-Maxwell equations with a null electromagnetic field \cite{VdB}. It must be remarked that there exists no analogue solution in higher dimensions \cite{OPZ}.

The space-time has three independent Killing vectors in general given by $\{\partial_x, \partial_y, y\partial_x-x\partial_y\}$, the corresponding isometry group acts transitively on spacelike 2-dimensional surfaces spanned by $\{\partial_x,\partial_y\}$. Thus, the third Killing vector is an isotropy. There is also a Kerr-Schild vector field \cite{CHS} $\vec\xi =\partial_ v$ satisfying
\be
\pounds_{\vec\xi} g_{\mu\nu} = \frac{2\dot m}{r}  k_\mu k_\nu \hspace{1cm} \pounds_{\vec\xi} k_\mu =0 \, . \label{ksvf}
\ee
This vector field is analogous to the Kodama vector field in spherically symmetric spacetimes \cite{K,BS}. The electromagnetic field inherits the symmetry $\{\partial_x,\partial_y\}$ but not the isotropy:
$$
\pounds_{y\partial_x-x\partial_y} \mathbf{F} =dv\wedge (h_ydx- h_x dy), \hspace{1cm} \pounds_{\vec\xi} \mathbf{F} = dv\wedge \left(\dot h_x dx + \dot h_y dy \right).
$$
Observe that the Kerr-Schild vector field is a symmetry of the electromagnetic field whenever $h_x,h_y$ are both constant, in which case the function $m(v)$ is linear in $v$.

The surfaces of transitivity, defined by constant values of $v$ and $r$, can describe (i) flat tori if one identifies $x \leftrightarrow x+a$ and $y\leftrightarrow y+b$, in which case they are compact with an area equal to $abr^2$; (ii) flat cylinders if  only one of the previous identifications is performed; and (iii) flat planes when $-\infty < x,y <\infty$. The mean curvature one-form \cite{Kriele,BS} for these surfaces is simply $\mathbf{H} = dr$ from where their two future null expansions can be easily extracted 
$$
\theta_1 =-\epsilon , \hspace{1cm} \theta_2 =-\epsilon \left(\frac{2m(v)}{r}+\frac{\Lambda}{3} r^2 \right) .
$$
Hence, the surfaces of transitivity are trapped if and only if $2m/r+\Lambda r^2/3 >0$. They are future- or past-trapped for $\epsilon =1$ or $-1$, respectively. Notice that the transitivity surfaces are always trapped for large enough $r$ if $\Lambda >0$ (de Sitter behavior at infinity), and for small enough $r$ if $m(v)>0$ independently of the sign of $\Lambda$. Actually, they are always trapped in the region with $m(v)>0$ if $\Lambda =0$. For $\Lambda <0$, the transitivity surfaces are trapped only in the region with $2m(v) > -\Lambda r^3/3 >0$ if this exists, and they can never be trapped for large enough $r$ (anti-de Sitter behavior). 

The hypersurface defined by
\be
{\cal H} : \hspace{3mm} 2m(v) +\frac{\Lambda}{3} r^3 =0
\ee 
(if this is feasible) is a marginally trapped tube, that is, a hypersurface foliated by marginally trapped surfaces. Observe that ${\cal H}$ exists for $\Lambda >0$, $=0$ or $<0$ only if $m(v)$ is negative, zero or positive, respectively, somewhere. One can easily compute the causal character of ${\cal H}$: it is non-spacelike if $\Lambda >0$ and actually timelike or null whenever $\dot m\neq 0$ or $\dot m=0$ respectively; non-timelike if $\Lambda <0$ with null portions wherever $\dot m=0$ and spacelike parts where $\dot m\neq 0$, in the last case these parts are dynamical horizons \cite{DFM}; if $\Lambda =0$ it is given by the null hypersurfaces $v=\hat v$ such that $m(\hat v)=0$.

There is a curvature singularity at $r\rightarrow 0$ unless $m(v)=0$ (in which case there is no electromagnetic radiation). This particular case with $m(v)=0$ has constant curvature $\Lambda/3$, so that the metric represents a region of de Sitter, flat, or anti-de Sitter space-time depending on the sign of $\Lambda$. In these cases $r=0$ is actually a horizon through which the metric is extendible.  Black holes in anti-de Sitter space-time obtained by identification along one symmetry generator were deeply analyzed in \cite{HP}. Other interesting particular cases arise if the electromagnetic radiation is absent ($h_x=h_y=0$) but we retain a non-vanishing constant $m(v)=M\neq 0$. These are metrics describing planar, cylindrical or toroidal black holes when $\Lambda <0$, and have been largely studied in the literature \cite{HL,L0,CZ,M,M1,V,KMV}. These cases, as follows from (\ref{ksvf}), have $\vec\xi$ as another Killing vector, so that they are stationary outside ${\cal H}$ which is a Killing horizon in this situation. One can check \cite{HL,L0,CZ,M,M1,V,KMV} that then $m(v)=M$ is proportional to the mass in the toroidal case, to mass per unit length in the cylindrical case, and to mass per unit area in the planar case. Therefore, negative values of $m(v)$ can be interpreted, at least when $\Lambda <0$, as negative values of the mass. Black holes with negative mass were discussed in \cite{M-}, but the remarkable thing about solution (\ref{ds2}) is that {\em the dominant energy condition holds everywhere}. This may be related to recent discussions on similar situations in de Sitter backgrounds \cite{BP,MP}. When $\Lambda =0$ but keeping $m(v)=M$ the metric can be seen to be isometric to the Kasner space-time \cite{Exact} with exponents $p_1=p_2=2/3,\,  p_3=-1/3$ if $M>0$, and isometric to the plane-symmetric Taub solution \cite{Exact} if $M<0$. 

The collapse to form non-spherical stationary black holes with a constant $m(v)=M$ and $\Lambda <0$ has been studied in several papers, such as for instance in \cite{SM} where the toroidal case treated herein was not explicitly considered but was later carried out in \cite{MNT}. In this reference \cite{MNT}, the collapse of perfect fluids describing anisotropic, as well as spatially inhomogeneous, interiors was fully described by matching these interiors to an exterior (\ref{ds2}) with constant $m(v)=M$. A collapse by matching was also considered in \cite{L}, where the dynamical metric (\ref{ds2}) was studied without realizing that the incoming flow of radiation is actually a null electromagnetic field, but this matching is incorrect.\footnote{In \cite{L} the metric (\ref{ds2}) is claimed to be matched to an interior Robertson-Walker metric for dust. However, this is impossible, as the Israel conditions \cite{I,MS} for a matching would require that the normal components of the energy momentum tensor be continuous at the matching hypersurface, and this cannot happen with a dust on one side and (\ref{emt}) on the other side.} 

The point I want to make in this sort Note is that the metric (\ref{ds2}) describes, appropriately, the generation of black holes by collapse of fully identified matter content: pure electromagnetic monochromatic waves with a well defined polarization. And there is no need for a matching procedure. Actually, this is just one situation of interest among a rich family of different behaviors that can be properly represented by (\ref{ds2}). Some outstanding cases are enumerated and briefly analyzed next.

\paragraph{1. Formation of a toroidal, cylindrical or planar black hole by sending light into an anti-de Sitter background}

By choosing $\Lambda <0$, $\epsilon =1$ and letting $h_x(v)=h_y(v)=0$ for all $v< v_0$ the function $m(v) =M$ is constant for all $v<v_0$. When $M$ is set to zero then the space-time is anti-de Sitter in this entire region. If light is then sent into the space-time by letting $h_x(v), h_y(v)$ to be non-zero in the interval $v_0\leq v \leq v_1$, a black hole enclosing a future singularity censored by an event horizon forms with a final constant $m(v)=M_f >0$ given by
\be
M_f =\frac{4\pi G}{c^4}\int^{v_1}_{v_0} [h_x^2(v)+h_y^2(v)]dv \, .\label{Mf0}
\ee

\begin{figure}[!ht]
\begin{center}
\includegraphics[height=10cm]{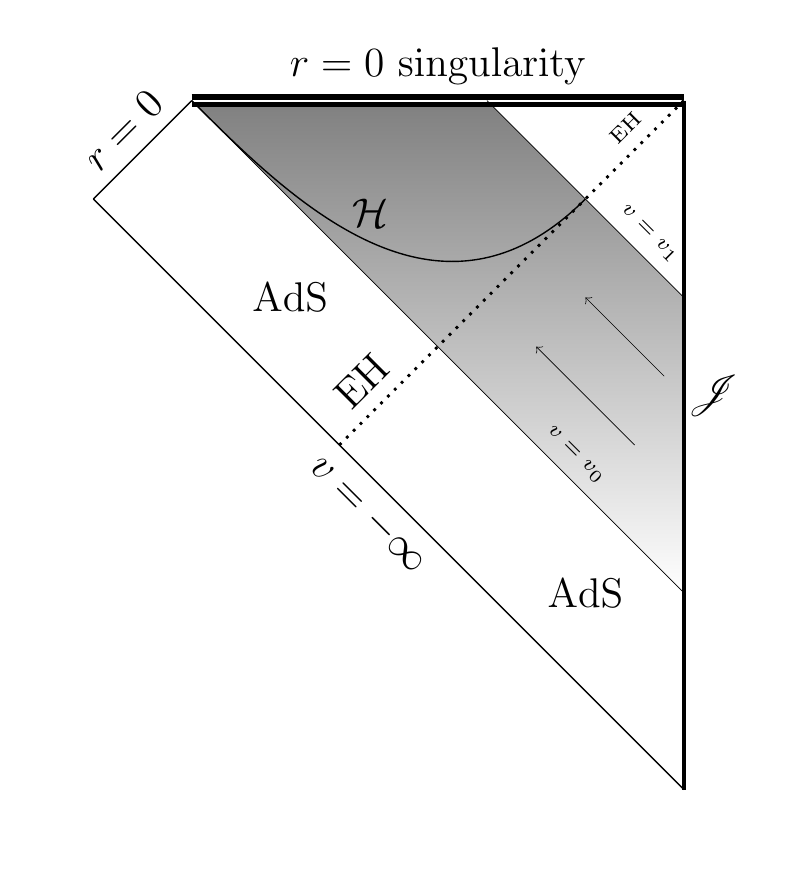}
\end{center}
\caption{\footnotesize{Schematic diagram of the formation of a toroidal (cylindrical or planar) black hole by sending light into an anti-de Sitter (AdS) background. As usual, null radial lines are drawn at 45$^o$ and the future direction is upwards. The spacetime is AdS until monochromatic light enters from $\scri$ (an infinity that is timelike)  at $v=v_0$ and flows along null hypersurfaces until it ceases at $v=v_1$. Thus, the shaded region contains a non-vanishing energy-momentum content due to the electromagnetic radiation solely. In the AdS part, the null hypersurfaces labeled as $r=0$ and $v=-\infty$ are horizons through which the space-time can be regularly extended. The spacetime has a final mass proportional to $M_f$ in (\ref{Mf0}). When the first photon reaches the $r=0$-horizon a dynamical horizon ${\cal H}$ develops that eventually (at $v=v_1$) merges with the event horizon EH of the black hole, which encloses a future spacelike singularity $r=0$.
Observe that EH starts developing in the AdS region.}}
\label{fig:BHformation}
\end{figure}

A Penrose-like diagram of this example is given in figure \ref{fig:BHformation}.

\paragraph{2. Transformation of a naked singularity into a black hole enclosing a clothed singularity by sending light into the former} When $M$ is not set to zero but rather is a {\em negative} constant in the previous situation, the space-time has a naked timelike singularity at $r=0$ for all $v\leq v_0$. Sending light again as before, and assuming that 
\be
M_f =\frac{4\pi G}{c^4}\int^{v_1}_{v_0} [h_x^2(v)+h_y^2(v)]dv +M \label{Mf}
\ee
\begin{figure}[!ht]
\begin{center}
\includegraphics[height=8cm]{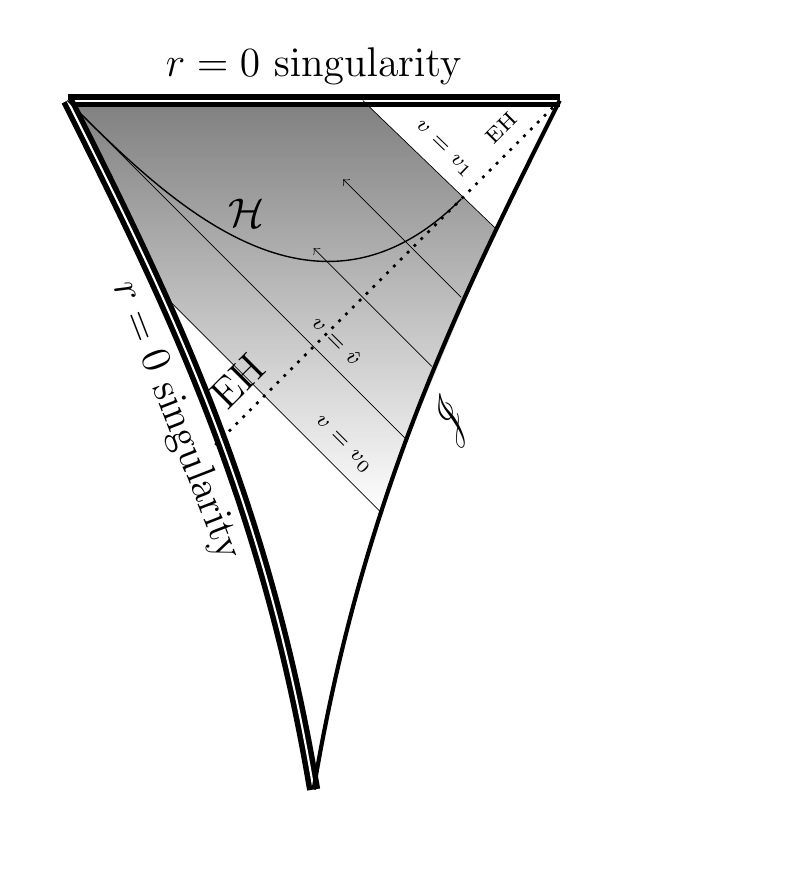}
\end{center}
\caption{\footnotesize{Schematic diagram of the transmutation of a naked singularity into a clothed one of a toroidal (cylindrical or planar) black hole by sending light into the former. The spacetime has a naked timelike singularity at $r=0$ with negative $m(v)=M$ until monochromatic light enters from  infinity $\scri$ at $v=v_0$ and flows along null hypersurfaces until it ceases at $v=v_1$, so the shaded region contains electromagnetic radiation solely. The spacetime has a final mass proportional to $M_f>0$ given in (\ref{Mf}) and therefore there is a value $v=\hat v$ where $m(\hat v)=0$. When the photons traveling along this hypersurface $v=\hat v$ reach the naked singularity ---as shown--- a dynamical horizon ${\cal H}$ develops that eventually (at $v=v_1$) merges with the event horizon EH of the black hole, enclosing a future spacelike singularity $r=0$.
}}
\label{fig:BH+naked+EH}
\end{figure}
is strictly positive, the singularity at $r=0$ transmutes into a spacelike one clothed by an event horizon which merges with the dynamical horizon ${\cal H}$. An illustrative diagram is presented in Figure \ref{fig:BH+naked+EH}. Similar situations arise for $\Lambda>0$ and $\Lambda =0$, but now without the formation of event horizons; in the former case there is a marginally trapped tube ${\cal H}$ which is partly null and partly timelike but not in the latter. Moreover, in the former case there is a past infinity $\scri^-$ which is spacelike while in the latter it is null.

\paragraph{3. Annihilation of a naked singularity by sending a fine-tuned amount of light}
\begin{figure}[!ht]
\begin{center}
\includegraphics[width=3.3cm]{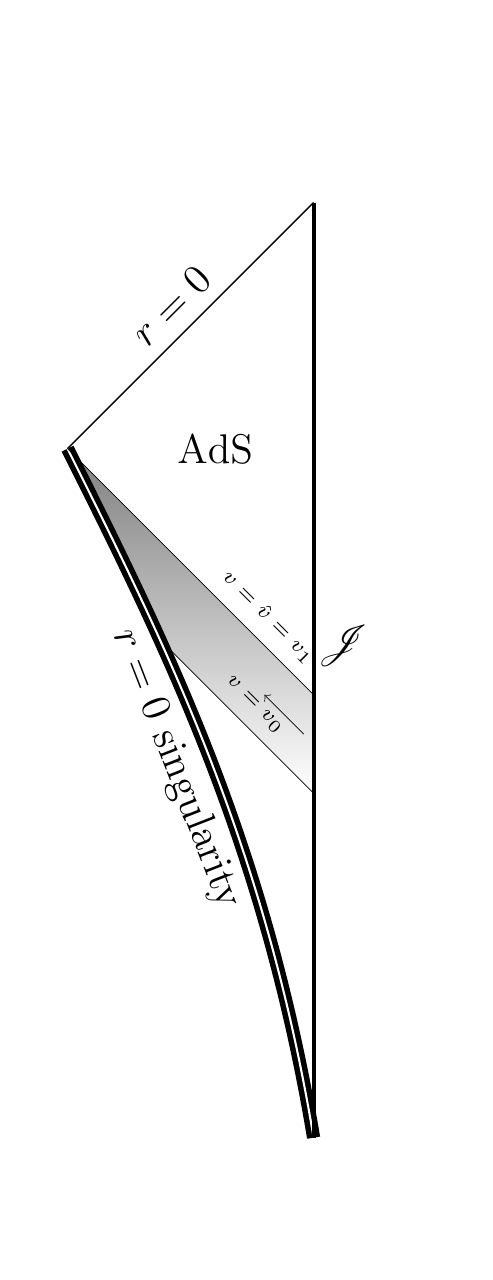}
\includegraphics[width=6.5cm]{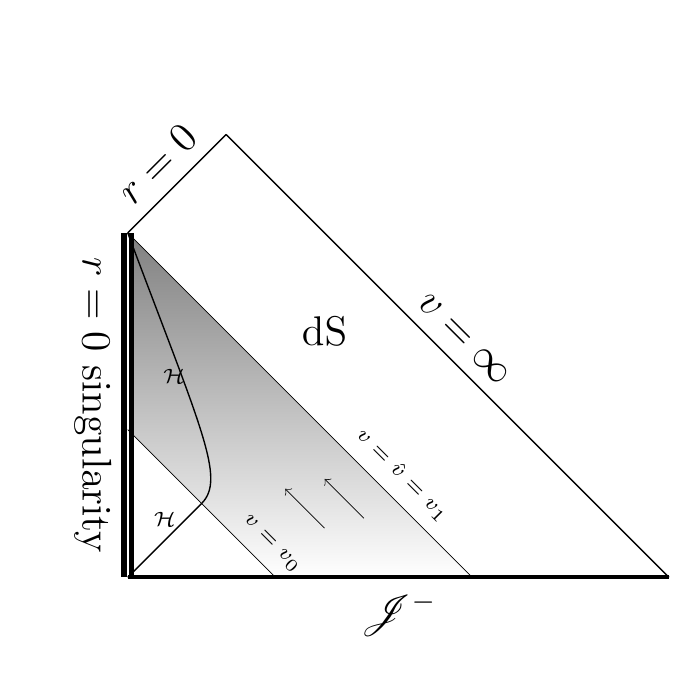}
\includegraphics[width=5.5cm]{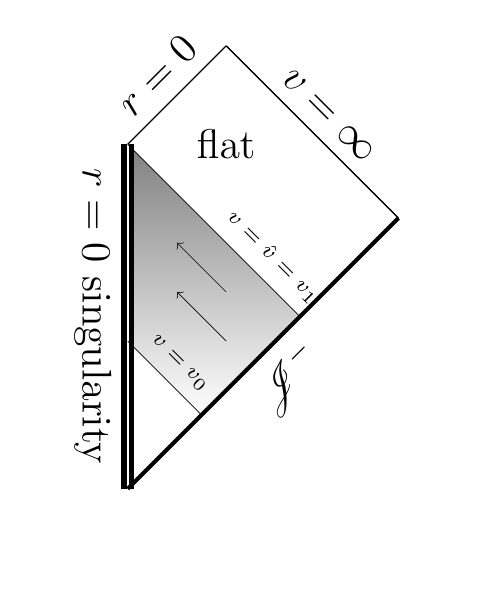}
\end{center}
\caption{\footnotesize{Diagrams of the complete annihilation of naked timelike singularities leading to a constant-curvature space-time. As in the previous figure, the spacetime has a naked timelike singularity at $r=0$ with negative $m(v)=M$ until monochromatic light enters from infinity $\scri$ at $v=v_0$ and stops flowing in at $v=v_1=\hat v$ with $m(\hat v)=0$, so the shaded regions contain only electromagnetic radiation. The spacetimes possess a final vanishing mass ($m(v)=0$ for all $v> \hat v$) and therefore they become anti-de Sitter, de Sitter or flat spacetimes according to whether $\Lambda$ is negative, positive or vanishing. These three cases are shown in that order from left to right. In the last two cases the null hypersurface marked by $v=\infty$ is a regular horizon and the metric can be extended beyond them. In all three cases the null hypersurfaces labeled as $r=0$ are horizons through which the space-time can be regularly extended too. A marginally trapped tube ${\cal H}$ which is partly null and then timelike exists only in the case with $\Lambda >0$, as shown.}}
\label{fig:Naked-destruction}
\end{figure}
Under the same assumptions as in the previous case, if $M<0$ and $h_x(v),h_y(v)$ are fine tuned together with $v_1-v_0$ such that the final $M_f$ in (\ref{Mf}) vanishes, the original timelike naked singularity simply disappears and the final outcome is a portion of anti-de Sitter , de Sitter, or flat space-time for negative, positive or vanishing $\Lambda$, respectively. The case with $\Lambda >0$ is the only one containing future trapped surfaces and a marginally trapped tube ${\cal H}$ ---which in this case is non-spacelike everywhere. Corresponding diagrams are given in figure \ref{fig:Naked-destruction}.

\paragraph{4. Time reversals} By setting $\epsilon =-1$ in (\ref{ds2}) one describes situations where the electromagnetic radiation is emitted outwardly towards the future; observe that now $\dot m\leq 0$. Then, for instance, models for the evaporation of a white hole by pure emission of light leading to anti-de Sitter space-time arise: this is simply the time reversal of {\em 1.} and the corresponding diagram is the same as in figure \ref{fig:BHformation} but turned upside down. 
Similarly, one can model the appearance of a naked timelike singularity in vacuum (with arbitrary $\Lambda$) by spontaneous emission of photons. Again, these are the time reversals of {\em 3.} and the corresponding diagrams are the same as in figure \ref{fig:Naked-destruction} interchanging future and past.

\section*{Acknowledgements}
Thanks to I. Bengtsson for bringing \cite{HP} to my attention.
Supported by grants
FIS2010-15492 (MICINN), GIU12/15 (Gobierno Vasco), P09-FQM-4496 (J. Andaluc\'{\i}a---FEDER) and UFI 11/55 (UPV/EHU).

\end{document}